# Waveguide-based single-pixel up-conversion infrared spectrometer

Qiang Zhang[1,2], Carsten Langrock[1], M. M. Fejer[1], Yoshihisa Yamamoto[1,2]

*1. Edward L. Ginzton Laboratory, Stanford University, Stanford, California 94305*
*2. National Institute of Informatics, 2-1-2 Hitotsubashi, Chiyoda-ku, Tokyo, 101-843, Japan*
*qiangzh@stanford.edu*

**Abstract:** A periodically poled lithium niobate (PPLN) waveguide-based single-pixel up-conversion infrared spectrometer was demonstrated. Sum-frequency generation between a 1.5-μm-band scanning pump laser and a 1.3-μm-band signal generated visible radiation which was detected by a silicon single-photon detector. The up-conversion spectrometer's sensitivity was two-orders-of-magnitude higher than that of a commercial optical spectrum analyzer.

___

**References and links**
1. B. H. Stuart, Infrared Spectroscopy: Fundamentals and Applications (Wiley, 2004).
2. Agilent, "Agilent 86140A Optical Spectrum Analyzer Family Technical Specifications," http://www.home.agilent.com/upload/cmc_upload/All/5968-1124E.pdf;
3. Princeton Instruments, "Spectroscopy Cameras OMAV," http://www.piacton.com/products/speccam/omav/default.aspx ;
4. E. J. Heilweil, "Ultrashort-pulse multichannel infrared spectroscopy using broadband frequency conversion in $LiO_3$," Opt. Lett. **14**, 551-553 (1989);
5. M. F. Decamp and A. Tokmakoff, "Up conversion multichannel infrared spectrometer," Opt. Lett. **30**, 1818-1820 (2005);
6. C. Langrock, E. Diamantini, R. V. Roussev, H. Takesue, Y. Yamamoto and M. M. Fejer, "Highly efficient single–photon detection at communication wavelengths by use of upconversion in reverse-proton-exchanged periodically poled $LiNbO_3$ waveguides," Opt. Lett. **30**, 1725-1727 (2005).
7. M. M. Fejer, G. A. Magel, D. H. Jundt, and R. L. Byer, "Quasi-Phase-Matched Second Harmonic Generation: Tuning and Tolerances," IEEE Journal of Quantum Electronics **28**, 2631–2654 (1992).
8. O. Kuzucu, F. N. C. Wong, S. Kurimura and S. Tovstonog, "Time-resolved characterization of single photons by upconversion," presented at the Conference on Lasers and Electro-Optics, San Jose, USA, 4-9, May, 2008.

___

## 1. Introduction

Near-infrared (NIR) frequency- and time-resolved spectroscopy are not only important in physical and chemical research, but also have many industrial applications, for example, optical communication, semiconductor microelectronics, and forensic analysis [1].

The optical spectrum analyzer (OSA) is a widely used NIR spectrometer, which is usually composed of a diffraction grating and an InGaAs linear array. Generally, OSAs cover a wide spectral range, allow for moderately fast wavelength sweeps, have a good spectral resolution, and don't require cryogenic cooling. For example, the Agilent 86140A [2] covers a wavelength range of several hundred nanometers, has a 40nm/50ms sweep rate, and a 0.025 nm resolution. However, due to the large dark current of the InGaAs array, the sensitivity of this instrument is -90 dBm at telecommunication wavelengths. Accoring to the manufacturer's definition of sensitivity [2], the instrument's noise equivalent power (NEP) is -98 dBm, which is not sufficient to capture single-photon-level spectra. Liquid-nitrogen-cooled InGaAs linear arrays can provide a smaller dark current and a lower NEP of -125 dBm [3], but the addition of cryogenic cooling reduces the usefulness for industrial applications.

One possible solution is to up convert NIR to visible light in a nonlinear medium via sum-frequency generation (SFG) [5] using a strong pump. The up-converted photons can then be detected using a linear silicon CCD array; silicon CCD arrays exhibit smaller dark currents than InGaAs arrays, which results in a better sensitivity.

Here, we present a waveguide-based single-pixel implementation of this method. In our implementation, we used a highly efficient periodically poled lithium niobate (PPLN) waveguide as the nonlinear medium [6]. Our spectrometer exhibited a two-orders-of-magnitudes lower NPE than the above mentioned commercial OSA. Furthermore, our spectrometer only used one single-photon detector instead of an array of detectors, which reduces cost and system complexity.

## 2. Experimental Implementation

In our experimental implementation, a NIR signal at 1.3 μm was combined with a strong pump at 1.55 μm using a fiber-based WDM before injecting both into a PPLN waveguide for sum-frequency generation (i.e. up conversion). The up-converted light was then detected by a single-photon counting module (SPCM). Since the periodically poled structure in the PPLN waveguide sets the quasi-phase-matching (QPM) condition for the SFG process, once the frequency of the pump light is set, the waveguide's acceptance bandwidth is determined by the QPM structure, effectively acting as a filter in the frequency domain [7]. Scanning the pump wavelength, and hence the center wavelength of this filter, allows us to trace out the spectrum of the signal light.

The resolution of the spectrometer is jointly determined by the QPM condition and the pump's spectral bandwidth, while the sweep rate is determined by the scanning speed of the pump laser and the data acquisition time. The spectral range of the up-conversion spectrometer is set by the spectral range of the pump laser and the QPM condition.

One of the advangtages of our spectrometer is its low NEP, expressed through detection efficiency (DE) and dark count (DC) rate. Compared to a bulk crystal, a PPLN waveguide can provide close to 100% internal signal conversion with low average pump power owing to its tight mode confinement over distances of several centimeters [6]. Therefore, the total DE of this type of spectrometer is limited by propagation, coupling, and reflection losses, as well as the SPCM's intrinsic quantum efficiency (QE).

The majority of dark counts of this system were caused by spurious nonlinear interactions inside the waveguide, including spontaneous anti-Stokes Raman scattering followed by up conversion. It has been shown that the dark counts increase quadratically with pump power [6]. Therefore, we could optimize the pump power to maximize the NEP of our spectrometer.

## 3. Experiment

As shown in Fig. 1, we used the light from a 1.3-μm Fabry-Perot laser diode as the signal for the spectral measurements. A C-band external-cavity tunable diode laser (ECDL) with a 1-pm linewidth, amplified by an erbium-doped fiber amplifier (EDFA), was used as the pump source. By adjusting the grating inside the ECDL, the emission spectrum could be scanned from 1535 nm to 1565 nm. Both the pump and the signal went through tunable attenuators and 20-dB optical splitters to control the pump and input signal power. They were then combined in a WDM before being input into the fiber-pigtailed PPLN waveguide device, which was operated at 100°C.

The PPLN waveguide used in our experiment was a 5-cm-long reverse-proton-exchange waveguide with a measured propagation loss of ~ 0.1 dB/cm at the pump wavelength. The passive signal power transmission through the non-AR-coated waveguide was 70%. We collected the SFG light with an AR-coated objective lens, and separated it from the pump and spurious light using a combination of short pass filter (Omega Optical SP-760), dichroic mirror (transmitting 1550 nm, reflecting 700 nm), prism, and spatial filter. The optical system's collection and transmission efficiency was 90%. The light was then focused onto the

SPCM with an AR-coated high-numerical-aperture lens. The quantum efficiency of the SPCM was ~ 70% at the SFG wavelength. The SPCM's output was connected to a computer-controlled counter, allowing synchronization with the tunable pump source.

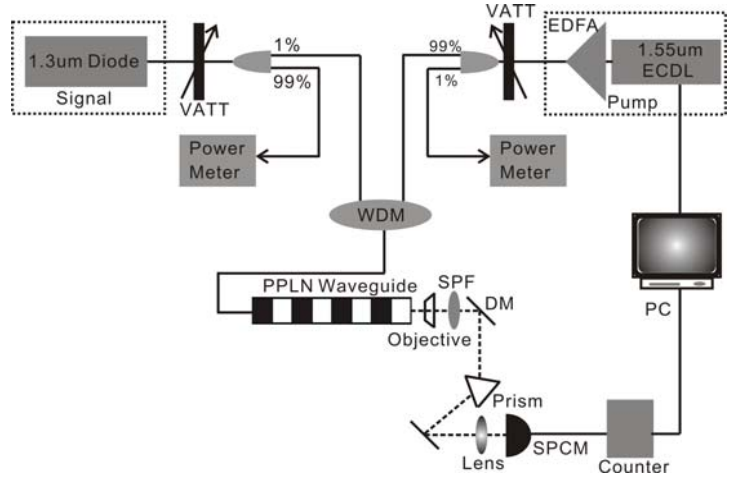

Fig. 1. Schematic of the waveguide-based single-pixel up-conversion spectrometer.

The QPM condition of our PPLN waveguide provided a 0.2-nm-wide acceptance bandwidth, much larger than the ECDL's 1-pm linewidth, effectively determining the spectral resolution of our up-conversion spectrometer. The maximum sweep rate of the ECDL was 100 nm/s, which was significantly slower than the 50 ns dead time of the SPCM and 10 ns dead time of the counter. Therefore the sweeping speed of our up conversion spectrometer can be as fast as 100 nm/s. However due to the communication speed between the PC and the counter in the experiment, we stepped the ECDL's wavelength by 0.05 nm every 100 ms.

The experimental results are shown in Figs. 2(a)-2(d). We first took a spectral measurement of the 1.3-μm laser diode using the commercial OSA (shown in Fig. 2(a)) when the output power of the laser diode was -0.067 dBm. Before injecting the signal into the up-conversion spectrometer, we set the input signal power to -98 dBm, corresponding to approximately 1 million photons per second, by adjusting the tunable attenuator. To optimize the NEP of the up-conversion spectrometer, we recorded the signal spectrum at various pump power levels by adjusting the EDFA's diode current. The results are shown in Figs. 2(b)-2(d). At a current setting of 275 mA, corresponding to a pump power of 9.86 dBm , the lowest NEP was achieved. At that pump power, the QE was around 5% and the root mean square of the DC was 550, resulting in an NEP of -118 dBm and a sensitivity of -110 dBm, two-orders-of-magnitude better than the OSA. To cover the spectrum of the 1.3-μm laser diode under test, the scan range of the ECDL was set to 1540 - 1565 nm. Its step size was set to 0.05 nm, smaller than the 0.2-nm resolution limit imposed by the length of the waveguide's QPM structure.

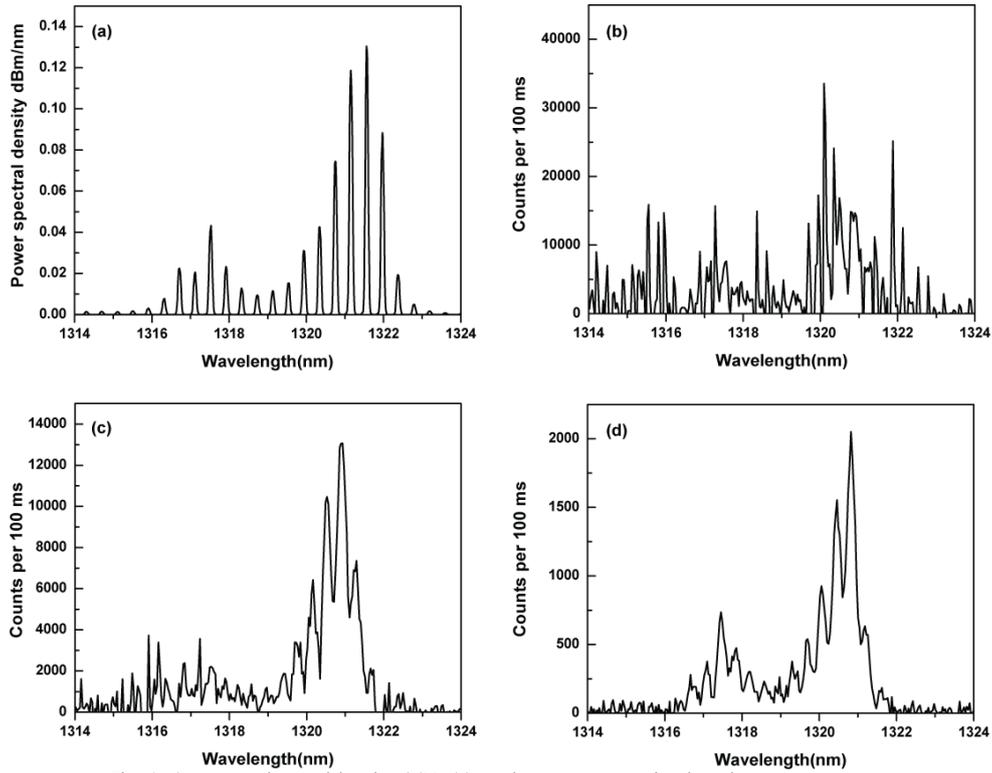

Fig. 2. Spectrum detected by the OSA (a), and our up-conversion-based spectrometer at different EDFA current settings, 550 mA (b), 450 mA (c) and 275 mA (d) using a 100-ms integration time.

With the optimized pump power, we measured the laser diode's spectrum at different signal levels to test the performance of the upconversion spectrometer. The results are shown in Figs. 3(a)-3(c). The highest peak of the spectrum could still be resolved down to an input power of -110 dBm, which could not be achieved by an OSA.

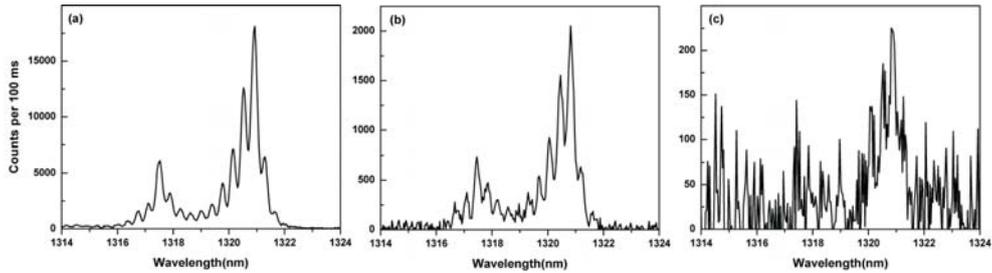

Fig. 3. The measured laser-diode spectrum at different input signal levels; -90 dBm (a), -100 dBm (b), -110 dBm (c). The SPCM was set to integrate for 100 ms.

## 4. Conclusion

We demonstrated a waveguide-based single-pixel up-conversion spectrometer, whose NEP is two-orders-of-magnitude lower than that of a conventional grating-based commercial OSA. Furthermore, this new type of spectrometer does not require cryogenic cooling and can be made very compact. The resolution of the spectrometer can be increased by using a longer QPM structure, which further increases the sensitivity of the setup. Adding AR coatings to the waveguide's end facets will increase the overall throughput by eliminating Fresnel losses. Due to its single-pixel, our new spectrometer maybe need less cost, but needs a sequential measurement, which is its disadvantage.

By using ultra-fast pump pulses instead of a CW laser, one can easily realize a time-resolved up-conversion spectrometer with the same setup. In fact, a bulk-crystal-based time-resolved up-conversion spectrometer has been reported recently [8].

**Acknowledgement**

QZ would like to thank Dr. Hiroki Takesue for useful discussions. This research was supported by NICT, the MURI center for photonic quantum information systems (ARO/ARDA program DAAD19-03-1-0199), SORST, CREST programs, Science and Technology Agency of Japan (JST), the U.S. Air Force Office of Scientific Research through contracts F49620-02-1-0240, the Disruptive Technology Office (DTO). We acknowledge the support of Crystal Technology, Inc.